# Symmetry-forbidden intervalley scattering by atomic defects in monolayer transition-metal dichalcogenides

Kristen Kaasbjerg,[1, *] Johannes H. J. Martiny,[1] Tony Low,[2] and Antti-Pekka Jauho[1]

[1]*Center for Nanostructured Graphene (CNG), Department of Micro- and Nanotechnology, Technical University of Denmark, DK-2800 Kongens Lyngby, Denmark*
[2]*Department of Electrical and Computer Engineering, University of Minnesota, Minneapolis, Minnesota 55455, USA*

Intervalley scattering by atomic defects in monolayer transition metal dichalcogenides (TMDs; $MX_2$) presents a serious obstacle for applications exploiting their unique valley-contrasting properties. Here, we show that the symmetry of the atomic defects can give rise to an unconventional protection mechanism against intervalley scattering in monolayer TMDs. The predicted defect-dependent selection rules for intervalley scattering can be verified via Fourier transform scanning tunneling spectroscopy (FT-STS), and provide a unique identification of, e.g., atomic vacancy defects ($M$ vs $X$). Our findings put the absence of the intervalley FT-STS peak in recent experiments in a different perspective.

***Introduction.***—Two-dimensional (2D) monolayers of transition metal dichalcogenides (TMDs; $MX_2$) are promising candidates for spin- and valleytronics applications [1]. Their hallmarks include unique valley-contrasting properties and strong spin-valley coupling [1, 2] exemplified by, e.g., valley-selective optical pumping [3–5], a valley-dependent Zeeman effect [6–9], and the valley Hall effect [10]. Such means to control the valley degree of freedom are instrumental for valleytronics applications.

Another prerequisite for a successful realization of valleytronics is a sufficiently long valley lifetime [11, 12]; atomic defects are a common limiting factor which can provide the required momentum for intervalley scattering due to their short-range nature. However, as illustrated in Fig. 1(a), the spin-orbit (SO) induced spin-valley coupling in the $K, K'$ valleys of 2D TMDs partially protects the valley degree of freedom against relaxation via intervalley scattering by nonmagnetic defects [2]. Due to the small spin-orbit splitting in the conduction band valleys [13, 14], only the valence-band valleys fully benefit from this protection. Identification of additional protection mechanisms in the conduction band would hence be advantageous for valleytronics in 2D TMDs.

In this work, we demonstrate that besides the spin-valley coupling, the symmetry and position of atomic defects give rise to unconventional selection rules for intervalley quasiparticle scattering in 2D TMDs. As illustrated in Fig. 1(b), we find that for defects with threefold rotational symmetry ($C_3$), e.g., atomic vacancies, intervalley $K \leftrightarrow K'$ scattering in the conduction band is forbidden for defects centered on the $X$ site while it is allowed for $M$ centered defects. In the valence band, intervalley scattering is forbidden in both cases. Analogous selection rules for the intervalley coupling due to confinement potentials in 2D TMD based quantum dots have previously been noted [15].

Our findings can be readily verified with scanning tunneling spectroscopy (STS) which has provided valuable insight to the electronic properties of 2D TMDs [16–20]. In particular, Fourier transform STS (FT-STS) is a powerful method for investigating atomic defects and their scattering properties in 2D materials [21, 22]. The measured STS map is a probe of the local density of states (LDOS) whose real-space modulation, resembling Friedel oscillations, originates from quasiparticle interference (QPI) between electronic waves scattered by defects. Hence, the Fourier transform of the STS map provides direct access to the available scattering channels in **q** space, and has shed important light on defect scattering in, e.g., graphene [23–30], monolayer TMDs [18, 19], and black phosphorus [31].

In the above-mentioned STS experiments on TMDs, the strong spin-valley coupling in the valence band of WSe$_2$ was confirmed by the missing $K \leftrightarrow K'$ intervalley peak in the FT-STS spectrum [18, 19]. Surprisingly, the intervalley peak was also missing in the conduction band where intervalley scattering should be allowed [18, 19] [see Fig. 1(a)].

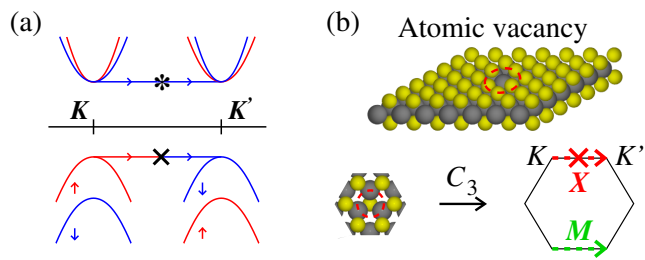

FIG. 1. Symmetry-dependent defect scattering in monolayer TMDs. (a) Sketch of the band structure near the $K, K'$ points. The strong spin-valley coupling in the valence band suppresses intervalley scattering (×). In the conduction band, the small spin-orbit splitting, in principle, allows for intervalley scattering (∗). However, for defects with threefold rotational symmetry ($C_3$), additional selection rules arise which protect against intervalley scattering. (b) Atomic sulfur vacancy in 2D $MoS_2$ showing the $C_3$ symmetry of the vacancy site. The vacancy-dependent selection rules for $K \leftrightarrow K'$ intervalley scattering in the conduction band are illustrated in the bottom part, showing that only $M$ vacancies produce intervalley scattering (green arrow). This allows for a unique identification of the vacancy type with FT-STS.

Here, we demonstrate the effect of symmetry on quasiparticle scattering by atomic vacancies which are among the most common types of defects in 2D TMDs [32–38]. For this purpose, we perform atomistic density-functional (DFT)-based $T$-matrix calculations [39] of FT-STS and QPI spectra for vacancies in two archetypal TMDs: the direct gap [44], small SO split MoS$_2$, and the indirect gap [17], large SO split [13, 14] WSe$_2$. As we show, the $K \leftrightarrow K'$ conduction-band intervalley FT-STS peak is strongly suppressed for $X$ vacancies while it appears clearly for $M$ vacancies, thus offering an appealing explanation for its conspicuous absence in experiments [18, 19]. Our findings furthermore show that FT-STS allows for a unique identification of the vacancy type, and indicate that the valley dynamics of carriers and excitons in 2D TMDs are not affected by disorder if $M$-type defects can be avoided.

***Symmetry-dependent intervalley scattering.***—We consider first the effect of symmetry on intervalley scattering by defects in 2D TMDs. The selection rules can be deduced within the framework of the low-energy Hamiltonian [2],

$$\mathcal{H}(\mathbf{k}) = at\left(\tau k_x \hat{\sigma}_x + k_y \hat{\sigma}_y\right) + \frac{\Delta}{2}\hat{\sigma}_z + \tau\lambda\frac{\hat{1}-\hat{\sigma}_z}{2}\hat{s}_z, \quad (1)$$

describing the band structure in the $K, K'$ valleys sketched in Fig. 1(a). Here, $a$ is the lattice constant, $t$ is a hopping parameter, $\tau = \pm 1$ is the $K, K'$ valley index, $\Delta$ is the band gap, $2\lambda$ is the SO splitting at the top of the valence band, and $\hat{\sigma}$, $\hat{\tau}$ and $\hat{s}$ are Pauli matrices in the symmetry-adapted spinor basis, valley and spin space, respectively. The symmetry-adapted basis is spanned by the $M$ $d$-orbitals $|\phi_{v\tau}\rangle = 1/\sqrt{2}\left(|d_{x^2-y^2}\rangle + i\tau|d_{xy}\rangle\right)$ and $|\phi_{c\tau}\rangle = |d_{z^2}\rangle$ which dominate the states in the valence ($v$) and conduction ($c$) bands, respectively [45, 46].

In 2D TMDs, defects such as atomic vacancies have $C_3$ symmetry, i.e. $\hat{V}_i = C_3\hat{V}_iC_3^\dagger$ where $\hat{V}_i$ is the scattering potential for defect type $i$ and $C_3$ is the operator for threefold rotations by $\pm 2\pi/3$ around the defect center. The intervalley matrix element ($\tau \neq \tau'$) between the high-symmetry $K, K'$ points can thus be written

$$\langle n\tau|\hat{V}_i|n\tau'\rangle = \langle n\tau|C_3^\dagger C_3\hat{V}_iC_3^\dagger C_3|n\tau'\rangle$$
$$= \langle n\tau|C_3^\dagger \hat{V}_i C_3|n\tau'\rangle \equiv \gamma_{i,n}^{\tau\tau'}\langle n\tau|\hat{V}_i|n\tau'\rangle, \quad (2)$$

where $n$ is the band index (including spin) and $\hat{I} = C_3^\dagger C_3$ is the identity operator. As $C_3$ belongs to the group of the wave vector at the $K, K'$ points ($C_{3h}$), the Bloch functions transform according to the irreducible representation of $C_{3h}$, $C_3|n\tau\rangle = w_{i,n\tau}|n\tau\rangle$ where $w_{i,n\tau}$ denotes the eigenvalues of $C_3$. The matrix element can thus be expressed in terms of the complex scalar $\gamma_{i,n}^{\tau\tau'} = w_{i,n\tau}^*w_{i,n\tau'}$ as indicated in the last equality of (2). Our analysis shows that $\gamma_{i,n}^{\tau\tau'} = 1$ only if the defect is centered on an $M$ site and $n = c$ [39]. In all other cases $\gamma_{i,n}^{\tau\tau'} \neq 1$, and the intervalley matrix element vanishes identically by virtue of Eq. (2).

The symmetry argument is completely general, and thus applies to all types of $M, X$-centered defects in 2D TMDs with $C_3$ symmetry, e.g., complex defect structures [32, 36], adatoms, and substitutional atoms [37]. As Eq. (1) is diagonal in spin space, it furthermore holds for intervalley spin-flip scattering by magnetic defects.

***FT-STS theory.***—Next, we outline a general $T$-matrix based Green's function approach for the calculation of the FT-STS spectra. In STS, the measured real-space QPI pattern is related to the differential conductance $dI/dV \propto \rho(\mathbf{r},\varepsilon)$ [47], and hence the LDOS $\rho(\mathbf{r},\varepsilon) = -1/\pi\text{Im}[G(\mathbf{r},\mathbf{r};\varepsilon)]$ where $G(\mathbf{r},\mathbf{r}';\varepsilon) = \langle\mathbf{r}|\hat{G}(\varepsilon)|\mathbf{r}'\rangle$ is the Green's function (GF) in real-space in the presence of a defect. Expressing the GF in a basis of Bloch states $\psi_{n\mathbf{k}}(\mathbf{r})$, $G(\mathbf{r},\mathbf{r}';\varepsilon) = \sum_{mn}\sum_{\mathbf{k}\mathbf{k}'}\psi_{n\mathbf{k}}^*(\mathbf{r})\psi_{m\mathbf{k}}(\mathbf{r}')G_{\mathbf{k}\mathbf{k}'}^{mn}(\varepsilon)$, where $\mathbf{k}$ is the wave vector and $m, n$ band indices, the FT-STS spectrum given by the 2D Fourier transform of $\rho(\mathbf{r},\varepsilon)$ can be obtained as [39]

$$\rho(\mathbf{q}+\mathbf{G},\varepsilon) = \int d\mathbf{r}\, e^{-i(\mathbf{q}+\mathbf{G})\cdot\mathbf{r}_\parallel}\rho(\mathbf{r},\varepsilon)$$
$$= \frac{1}{2\pi i}\sum_{mn,\mathbf{k}}n_{\mathbf{k},\mathbf{q}}^{mn}(\mathbf{G})\left[G_{\mathbf{k},\mathbf{k}+\mathbf{q}}^{mn}(\varepsilon)^* - G_{\mathbf{k}+\mathbf{q},\mathbf{k}}^{nm}(\varepsilon)\right], \quad (3)$$

where $\mathbf{r} = (\mathbf{r}_\parallel, z)$, $\mathbf{k}, \mathbf{q} \in$ 1st Brillouin zone (BZ), $\mathbf{G}$ is a reciprocal lattice vector, and $G_{\mathbf{k}\mathbf{k}'}^{mn}(\varepsilon) = \langle\psi_{m\mathbf{k}}|\hat{G}(\varepsilon)|\psi_{n\mathbf{k}'}\rangle$ is the Bloch function representation of the GF. The matrix element $n_{\mathbf{k},\mathbf{q}}^{mn}(\mathbf{G}) = \langle\psi_{m\mathbf{k}}|e^{-i(\mathbf{q}+\mathbf{G})\cdot\hat{\mathbf{r}}_\parallel}|\psi_{n\mathbf{k}+\mathbf{q}}\rangle$ is important in many aspects. For example, it describes the FT-STS Bragg peaks ($\mathbf{G} \neq \mathbf{0}$), and hence the atomic modulation of the LDOS inside the unit cell. It also plays a central role in systems with (pseudo) spin texture, e.g., graphene and spin-orbit materials, as it contains the spinor overlap [48]. This is less important in 2D TMDs where the eigenstates of Eq. (1) are characterized by predominantly polarized spinor states [49] with trivial pseudospin, $\hat{\sigma}$, and spin, $\hat{s}$, textures.

For a single defect, the *exact* GF taking into account multiple scattering off the defect is given by the $T$ matrix as

$$\mathbf{G}_{\mathbf{k}\mathbf{k}'}(\varepsilon) = \delta_{\mathbf{k},\mathbf{k}'}\mathbf{G}_\mathbf{k}^0(\varepsilon) + \mathbf{G}_\mathbf{k}^0(\varepsilon)\mathbf{T}_{\mathbf{k}\mathbf{k}'}(\varepsilon)\mathbf{G}_{\mathbf{k}'}^0(\varepsilon), \quad (4)$$

where the boldface symbols denote matrices in band and spin indices, and the diagonal bare GF is given by the band energies, $G_{0\mathbf{k}}^0(\varepsilon) = (\varepsilon - \varepsilon_{n\mathbf{k}} + i\eta)^{-1}$. The last term in Eq. (4) comprises the nondiagonal, defect-induced correction $\delta\mathbf{G}_{\mathbf{k},\mathbf{k}+\mathbf{q}}$ to the GF. To isolate the FT-STS features related to the defect, we substitute $G \to \delta G$ in Eq. (3) in our FT-STS calculations.

The $T$ matrix obeys the integral equation

$$\mathbf{T}_{\mathbf{k}\mathbf{k}'}(\varepsilon) = \mathbf{V}^i_{\mathbf{k}\mathbf{k}'} + \sum_{\mathbf{k}''} \mathbf{V}^i_{\mathbf{k}\mathbf{k}''} \mathbf{G}^0_{\mathbf{k}''}(\varepsilon) \mathbf{T}_{\mathbf{k}''\mathbf{k}'}(\varepsilon), \quad (5)$$

where $V^{mn}_{i,\mathbf{k}\mathbf{k}'}$ are matrix elements of the defect potential and the second term describes virtual transitions to intermediate states with wave vector $\mathbf{k}''$.

For nonmagnetic defects, we take $\hat{V}_i = V_i(\hat{\mathbf{r}}) \otimes \hat{s}_0$ where $\hat{s}_0$ is the identity operator in spin space. With the spin indices written out explicitly, the defect matrix elements can be expressed as

$$V^{mn}_{i,\mathbf{k}\mathbf{k}'}(s,s') = \langle m\mathbf{k}s|\hat{V}_i|n\mathbf{k}'s'\rangle$$
$$= \sum_{s_z} \langle m\mathbf{k}s; s_z|V_i(\hat{\mathbf{r}})|n\mathbf{k}'s'; s_z\rangle, \quad (6)$$

with $|\cdot; s_z\rangle$ denoting the $s_z = \pm 1$ spinor component of the wave function. Here, we use a DFT method based on an atomic supercell model for the defect site illustrated in Fig. 1(b) to calculate the defect matrix elements [39].

As an example, Fig. 2 shows the spin-diagonal conduction-band matrix elements for Mo and S vacancies in 2D MoS$_2$. While the Mo vacancy gives rise to intravalley (short arrow) and intervalley (long arrow) couplings, the intervalley matrix element for the S vacancy vanishes, thus confirming the symmetry-based predictions in Eq. (2). Furthermore, we note that the matrix element in the $K, K'$ valleys is an order of magnitude larger for Mo than for S vacancies. In a simple picture where only $K, K'$ intra- and intervalley scattering with a constant matrix element $V_0$ is considered, the $T$ matrix becomes $T(\varepsilon) = V_0/[1 - gV_0\bar{G}_0(\varepsilon)]$, where $\bar{G}_0(\varepsilon) = \int \frac{d\mathbf{k}}{(2\pi)^2} G^0_{c\mathbf{k}}(\varepsilon) \propto \rho_c$, $\rho_c \approx 0.01$ eV$^{-1}$ Å$^{-2}$ is the density of states, and the valley multiplication factor $g = 2$ ($= 1$) for $M$ ($X$; only intravalley scattering) vacancies. Together with the values for $V_0$ extracted from Fig. 2, this allows us to identify $M$ ($g\rho_c V_0 > 1$) and $X$ ($g\rho_c V_0 < 1$) vacancies as *strong* (unitary), $T(\varepsilon) \approx -1/g\bar{G}_0(\varepsilon)$, and *weak*, $T(\varepsilon) \approx V_0$, defects, respectively.

The FT-STS calculations presented below are based on full BZ $\mathbf{k}, \mathbf{q}$-point samplings of the band structures, defect matrix elements, and $n^{mn}_{\mathbf{k},\mathbf{q}}(\mathbf{G})$ matrix elements, all obtained with DFT-LDA including SO interaction [39]. Our approach naturally goes beyond the low-energy description in Eq. (1), which is essential as both the $K$ and $Q$ valleys are relevant for quasiparticle scattering in 2D TMDs. As intervalley scattering in the valence band is suppressed by (i) the large spin-valley coupling, and (ii) the $C_3$ symmetry of the vacancies, the valence-band FT-STS spectra are rather simple [18, 19], and we here limit the discussion to the conduction band. We furthermore focus on features related to the symmetry-forbidden intervalley scattering defering a complete analysis to a forthcoming paper.

***FT-STS and QPI spectra.***—The calculated band structures and FT-STS spectra for atomic vacancies in MoS$_2$ and WSe$_2$ are summarized in Fig. 3. The different conduction-band structures in the two materials ($K$ vs $Q$ valley alignment and magnitude of the SO splitting) shown in the insets in Fig. 3(a) and the vacancy-dependent intervalley matrix element, result in markedly different spectra between the materials as well as the vacancy type.

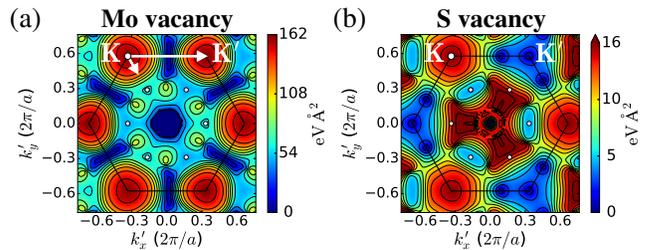

FIG. 2. Defect matrix elements for the conduction band in 2D MoS$_2$ calculated with our DFT-based supercell method. The plots show $|V^{cc}_{i,\mathbf{k}\mathbf{k}'}(s,s)|$ for (a) a Mo, and (b) a S vacancy as a function of $\mathbf{k}'$ with the initial state fixed to $\mathbf{k} = \mathbf{K}$. Note the different disorder strengths (colorbar scales) for the two types of vacancies as well as the vanishing intervalley matrix element [long arrow in (a)] for S vacancies.

In general, the FT-STS spectra close to the band edge ($\varepsilon \approx 0$; see Ref. [39]) are characterized by featureless spots at the points in $\mathbf{q}$ space corresponding to intravalley ($\mathbf{q} = \mathbf{0}$) and intervalley scattering [$\mathbf{q}_{1-5}$ in Fig. 3(b)]. The spot intensities are governed by the $T$-matrix scattering amplitude and valley degeneracy. For the Bragg peaks, the intensity is reduced compared to those in the first BZ due to the phase-factor matrix element $n^{mn}_{\mathbf{k},\mathbf{q}}(\mathbf{G})$.

In MoS$_2$ the SO splitting in the conduction band is small, $\sim 3$ meV, thereby allowing for spin-conserving $K \leftrightarrow K'$ intervalley scattering ($\mathbf{q}_{1,2}$) near the band edge. Hence, intervalley peaks at $\mathbf{q} = \mathbf{K}, \mathbf{K}'$ are to be expected. In WSe$_2$ the $Q$ valley is lower than the $K$ valley and the SO splitting is much larger ($\sim 250$ meV in the $Q$ valley and $\sim 50$ meV the $K$ valley), hence a $\mathbf{q} \approx \mathbf{M}$ peak due to $Q \leftrightarrow Q'$ intervalley processes ($\mathbf{q}_3$) will appear instead.

The above is indeed the case in the FT-STS spectra for $M$ vacancies shown in Fig. 3(c) for an energy $\varepsilon = 75$ meV above the band edge [dashed lines in the insets in Fig. 3(a)]. At this energy, the spots have developed into features (see the zoomed insets) which are dominated by processes involving nesting vectors between parallel segments of the constant energy contour being probed. In MoS$_2$ with almost isotropic energy contours, $\varepsilon(k) = \varepsilon$, intravalley backscattering with $q = 2k$ therefore produces circular features. Trigonal warping of the constant energy surfaces gives rise to additional approximate nesting vectors which produce starlike patterns with hexagonal symmetry around the $\Gamma$ point and triangular symmetry near the $K, K'$ points as in graphene [30]. The intervalley features are weaker than the intravalley feature because intravalley processes in the $K$ and $K'$ valleys add up, while the two $K \leftrightarrow K'$ intervalley processes have distinct wave vectors, $\mathbf{q} \approx \pm\mathbf{K}$. In $WSe_2$, both the $Q$ and $K$ valleys are accessible at $\varepsilon = 75$ meV, and therefore intervalley features around $\mathbf{q} \approx \mathbf{M}$, $\mathbf{q} \approx \mathbf{K}$ as well as $\mathbf{q} \approx \mathbf{Q}$ are observed. They are associated with $Q \leftrightarrow Q/K \leftrightarrow Q$ ($\mathbf{q}_{3/4}$), $K \leftrightarrow K'$ ($\mathbf{q}_{1,2}$), and $K \leftrightarrow Q$ ($\mathbf{q}_5$) processes, respectively. The central intravalley feature in $WSe_2$ has more structure than in $MoS_2$ as it has contributions from both $K$ and $Q$ intravalley processes.

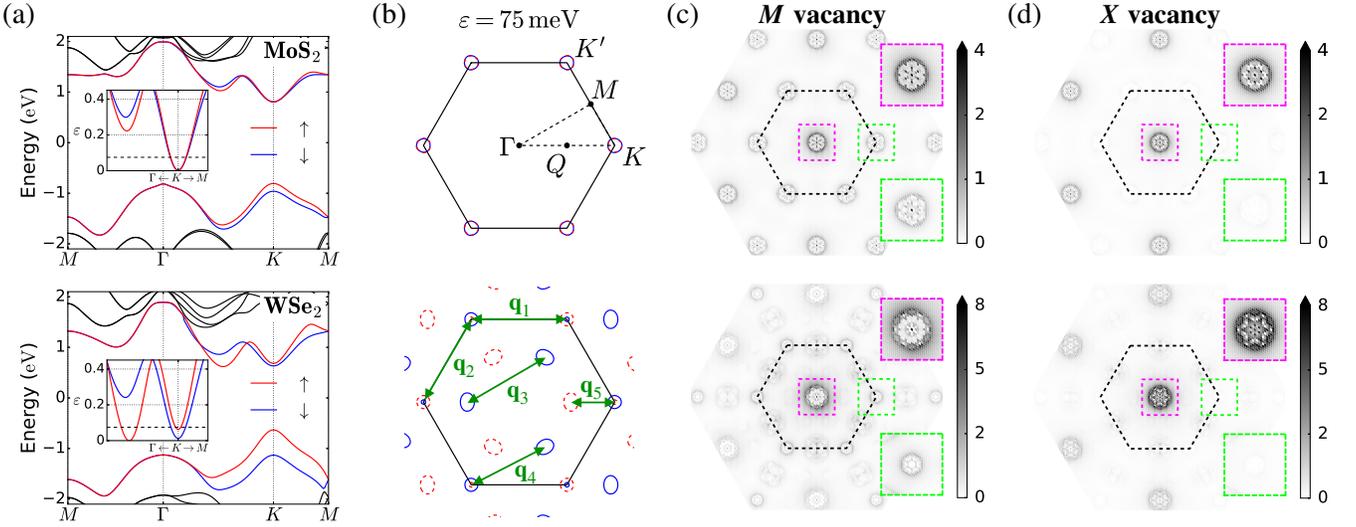

FIG. 3. Band structures and FT-STS spectra for atomic vacancies in $MoS_2$ (top) and $WSe_2$ (bottom). (a) Band structures including SO interaction. The insets show a zoom of the SO split conduction-band $K, Q$ valleys with the energy $\varepsilon = E - E_c$ measured relative to the band edge $E_c$. The dashed lines indicate the energy of the constant-energy surfaces in (b) and the FT-STS spectra in (c),(d). (b) Constant-energy surfaces in $\mathbf{k}$ space for $\varepsilon = 75$ meV, together with high-symmetry $\mathbf{k}$ points in the Brillouin zone (top) and representative intervalley $\mathbf{q}$ vectors (bottom). (c), (d) FT-STS spectra at $\varepsilon = 75$ meV for (c) $M =$ Mo, W and (d) $X =$ S, Se vacancies. The boxes show zooms of the marked regions.

At even higher energies (not shown), the $K$ and $Q$ valleys are available in both $MoS_2$ and $WSe_2$, and the FT-STS spectra become highly complex.

In contrast to the FT-STS spectra for $M$ vacancies, the spectra for $X$ vacancies in Fig. 3(d) show that the anticipated intervalley feature at $\mathbf{q} \approx \mathbf{K}$ ($\mathbf{q}_{1,2}$) is strongly suppressed for both $MoS_2$ and $WSe_2$. This is a direct consequence of the symmetry-forbidden $K \leftrightarrow K'$ intervalley matrix element which suppresses intervalley scattering also in the vicinity of the high-symmetry $K, K'$ points [see Fig. 2(b)]. In $WSe_2$, also the $Q \leftrightarrow Q'$ ($\mathbf{q}_3$) and $Q \leftrightarrow K$ ($\mathbf{q}_{4,5}$) intervalley features are much weaker for $X$ vacancies, which can be traced back to overall small intervalley matrix elements.

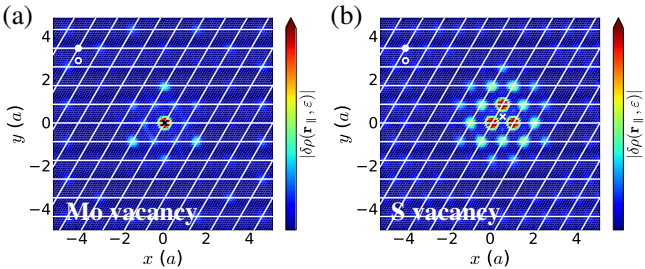

FIG. 4. Real-space QPI maps for 2D $MoS_2$ showing the defect-induced change in the LDOS $\delta\rho(\mathbf{r}_\parallel, \varepsilon)$ around (a) a Mo, and (b) a S vacancy. The lines show the unit cells of the lattice with lattice constant $a$, and the atomic positions inside the unit cell and the position of the vacancy are indicated by the symbols (solid circle: Mo; open circle: S; cross: vacancy).

The suppression of $K \leftrightarrow K'$ intervalley scattering for $X$ vacancies leaves a clear fingerprint in the real-space LDOS as demonstrated by the QPI maps in Fig. 4 for Mo and S vacancies in $MoS_2$. They have been obtained by Fourier transforming the FT-STS spectra in Figs. 3(c) and 3(d), $\rho(\mathbf{r}_\parallel, \varepsilon) = \sum_\mathbf{G} \int \frac{d\mathbf{q}}{(2\pi)^2} e^{i(\mathbf{q}+\mathbf{G})\cdot \mathbf{r}_\parallel} \rho(\mathbf{q}+\mathbf{G}, \varepsilon)$. For both vacancies, the LDOS modulation has a threefold symmetry and decays away the vacancy site (marked by crosses). The observed atomic resolution can be attributed to the FT-STS Bragg peaks, and shows that the LDOS modulation is concentrated on the Mo sites of the lattice, in accordance with the Mo $d$-orbital character of the conduction-band states in the $K, K'$ valleys [cf. Eq. (1)]. Noticeably, the QPI map for the S vacancy stands out by the absence of an intervalley-scattering-induced cell-to-cell modulation of the LDOS in the vicinity of the vacancy, which is clearly visible for the Mo vacancy. At larger distances from the vacancy site, a slower modulation with wave length $2\pi/q$ ($\approx 10\,a$ at $\varepsilon = 75$ meV) due to intravalley backscattering, $q = 2k$, emerges.

***Conclusions and outlook.***—In conclusion, we have demonstrated (i) an unconventional symmetry-induced protection against intervalley scattering by atomic defects in 2D TMDs, and (ii) its fingerprint in conduction-band FT-STS spectra which allows for a unique identification of, e.g., the vacancy type. Our findings may offer an explanation why the $K \leftrightarrow K'$ intervalley FT-STS peak has not been observed in experiments [18, 19], and are also relevant for FT-STS on metallic TMDs [50].

We are convinced that our work in conjunction with further experimental FT-STS studies can provide a complete understanding of defect scattering in 2D TMDs. In addition, FT-STS may shed important light on band-structure issues in 2D TMDs, such as the magnitude of SO splittings [18], the $K$,$Q$-valley ordering in the conduction band which is sensitive to the SO strength [13, 14], and the subband structure and valley ordering in few-layer TMDs [51, 52]. Besides our reported FT-STS signatures, the suppression of intervalley scattering is expected to have implications for a wide range of effects in disordered 2D TMDs, e.g., the optical conductivity [53], magnetotransport [54–58], the valley Hall effect [59], Elliot-Yafet spin relaxation [60], and disorder-induced valley pumping [61].

***Acknowledgements.***—We would like to thank an anonymous reviewer on a related work [62] for suggesting we investigate observable implications of our symmetry finding for intervalley scattering. K.K. acknowledges support from the European Union's Horizon 2020 research and innovation programme under the Marie Sklodowska-Curie Grant Agreement No. 713683 (COFUNDfellows-DTU). T.L acknowledges support from the National Science Foundation under Grant No. NSF/EFRI-1741660. The Center for Nanostructured Graphene (CNG) is sponsored by the Danish National Research Foundation, Project No. DNRF103.

# Supplemental material for "Symmetry-forbidden intervalley scattering by defects in monolayer transition-metal dichalcogenides"

Kristen Kaasbjerg,[1, *] Johannes H. J. Martiny,[1] Tony Low,[2] and Antti-Pekka Jauho[1]

[1] *Center for Nanostructured Graphene (CNG), Dept. of Micro- and Nanotechnology, Technical University of Denmark, DK-2800 Kongens Lyngby, Denmark*
[2] *Department of Electrical and Computer Engineering, University of Minnesota, Minneapolis, MN 55455, USA*
(Dated: December 4, 2017)

## S1. THEORETICAL FT-STS CALCULATIONS

Under the assumption that the density of states of the STM tip varies slowly with energy, the $dI/dV$ characteristics at position $\mathbf{r}$ and voltage $eV = \varepsilon$ is proportional to the local density of states $\rho(\mathbf{r}, \varepsilon)$ of the sample[1],

$$\frac{dI(\mathbf{r}, \varepsilon)}{dV} \propto \rho(\mathbf{r}, \varepsilon) = -\frac{1}{2\pi i} \left[ G(\mathbf{r}, \mathbf{r}; \varepsilon) - G^*(\mathbf{r}, \mathbf{r}; \varepsilon) \right], \tag{S1}$$

where $G(\mathbf{r}, \mathbf{r}'; \varepsilon) = \langle \mathbf{r}|\hat{G}(\varepsilon)|\mathbf{r}'\rangle$, $\hat{G}(\varepsilon) = [\varepsilon - \hat{H} + i\eta]^{-1}$, is the real-space Green's function (GF) for a defect in the 2D material.

For a numerical evalutation of the FT-STS spectrum, it is convenient to express the GF in terms of Bloch states $\psi_{m\mathbf{k}}$ of the pristine lattice. By inserting the identity $I = \sum_{m\mathbf{k}} |\psi_{m\mathbf{k}}\rangle\langle\psi_{m\mathbf{k}}|$, the GF can be written as

$$G(\mathbf{r}, \mathbf{r}'; \varepsilon) = \langle \mathbf{r}|\hat{G}(\varepsilon)|\mathbf{r}'\rangle = \sum_{mn} \sum_{\mathbf{k}\mathbf{k}'} \langle \mathbf{r}|\psi_{m\mathbf{k}}\rangle \langle \psi_{m\mathbf{k}}|\hat{G}(\varepsilon)|\psi_{n\mathbf{k}'}\rangle \langle \psi_{n\mathbf{k}'}|\mathbf{r}'\rangle$$

$$= \sum_{mn} \sum_{\mathbf{k}\mathbf{k}'} \psi_{m\mathbf{k}}(\mathbf{r}) \psi_{n\mathbf{k}'}^*(\mathbf{r}') G_{\mathbf{k}\mathbf{k}'}^{mn}(\varepsilon), \tag{S2}$$

where $G_{\mathbf{k}\mathbf{k}'}^{mn}(\varepsilon) = \langle \psi_{m\mathbf{k}}|\hat{G}(\varepsilon)|\psi_{n\mathbf{k}'}\rangle$ is its Bloch function representation and the $\mathbf{k}, \mathbf{k}'$ sums are over the first Brillouin zone (BZ), here sampled with a discrete, equidistant $N_k \times N_k$ $\mathbf{k}$-point grid as illustrated in Fig. S1.

The FT-STS spectrum is given by the 2D Fourier transform (FT) of the LDOS in Eq. (S1). Here, we consider the $z$-integrated LDOS, which is a reasonable approach for 2D materials. Plugging in the Bloch function expansion of the GF and setting $\mathbf{r} = (\mathbf{r}_\parallel, z)$ where $\mathbf{r}_\parallel$ is the inplane component of the position, the $z$-integrated FT becomes

$$\rho(\mathbf{q} + \mathbf{G}, \varepsilon) = \int d\mathbf{r}\, e^{-i(\mathbf{q}+\mathbf{G}) \cdot \mathbf{r}_\parallel} \rho(\mathbf{r}, \varepsilon)$$

$$= -\frac{1}{2\pi i} \int d\mathbf{r}\, e^{-i(\mathbf{q}+\mathbf{G}) \cdot \mathbf{r}_\parallel} \left[ G(\mathbf{r}, \mathbf{r}; \varepsilon) - G^*(\mathbf{r}, \mathbf{r}; \varepsilon) \right]$$

$$= -\frac{1}{2\pi i} \sum_{mn} \sum_{\mathbf{k}\mathbf{k}'} \underbrace{\int d\mathbf{r}\, \psi_{m\mathbf{k}}^*(\mathbf{r}) e^{-i(\mathbf{q}+\mathbf{G}) \cdot \mathbf{r}_\parallel} \psi_{n\mathbf{k}'}(\mathbf{r}')}_{\delta_{\mathbf{k}', \mathbf{k}+\mathbf{q}} n_{\mathbf{k},\mathbf{q}}^{mn}(\mathbf{G})} \left[ G_{\mathbf{k}',\mathbf{k}}^{nm}(\varepsilon) - G_{\mathbf{k},\mathbf{k}'}^{mn}(\varepsilon)^* \right]$$

$$= \frac{1}{2\pi i} \sum_{mn} \sum_{\mathbf{k}} n_{\mathbf{k},\mathbf{q}}^{mn}(\mathbf{G}) \times \left[ G_{\mathbf{k},\mathbf{k}+\mathbf{q}}^{mn}(\varepsilon)^* - G_{\mathbf{k}+\mathbf{q},\mathbf{k}}^{nm}(\varepsilon) \right], \tag{S3}$$

where $\mathbf{q} \in$ 1. BZ, $\mathbf{G}$ is a reciprocal lattice vector, and $n_{\mathbf{k},\mathbf{q}}^{mn}(\mathbf{G}) = \langle \psi_{m\mathbf{k}}|e^{-i(\mathbf{q}+\mathbf{G})\cdot\hat{\mathbf{r}}_\parallel}|\psi_{n\mathbf{k}+\mathbf{q}}\rangle = \langle u_{m\mathbf{k}}|e^{-i\mathbf{G}\cdot\hat{\mathbf{r}}_\parallel}|u_{n\mathbf{k}+\mathbf{q}}\rangle$, where $u_{m\mathbf{k}}$ is the periodic part of the Bloch functions, is a phase-factor matrix element. The latter is important in both technical and practical aspects. In numerical calculations, it cancels the arbitrary phase on the wave functions in $G_{\mathbf{k}\mathbf{k}'}^{mn}(\varepsilon)$, thus leaving the expression (S3) gauge invarient as it should be. In the FT-STS spectra, it may affect the structure of the intra- and intervalley features, and is the reason that the Bragg peaks, $\mathbf{G} \neq \mathbf{0}$, in general, must be expected to differ from the corresponding $\mathbf{G} = \mathbf{0}$ peaks inside the first BZ.

All results presented in the main manuscript are based on the general expression for the FT-STS spectrum in Eq. (S3). However, we note that simpler variants which follow from this general expression are often encountered in the literature. For example, if we disregarding the reciprocal lattice vector and assume that the periodic parts of the Bloch functions are orthogonal, i.e. $\langle u_{m\mathbf{k}}|u_{n\mathbf{k}+\mathbf{q}}\rangle = \delta_{mn}$, the expression reduces to

$$\rho(\mathbf{q}, \varepsilon) \approx \frac{1}{2\pi i} \sum_{\mathbf{k}} \text{Tr}\left[ \mathbf{G}_{\mathbf{k},\mathbf{k}+\mathbf{q}}(\varepsilon)^* - \mathbf{G}_{\mathbf{k}+\mathbf{q},\mathbf{k}}(\varepsilon) \right]. \tag{S4}$$

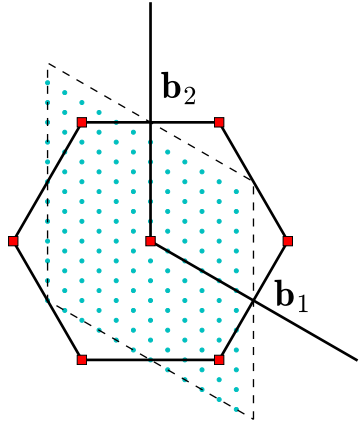

FIG. S1. Equidistant $N_k \times N_k$ grid used for the $\mathbf{k},\mathbf{q}$-point samplings of the rhombic BZ in the numerical calculation of the defect GF and the FT-STS spectra. The plot shows a $12 \times 12$ BZ grid while a $75 \times 75$ grid was used in the actual calculations presented in this work.

Depending on the approxiation used for the defect GF, e.g., the Born approximation, this may be simplified further. Note that since $\rho(\mathbf{r})$ is real-valued, it follows that $\rho(\mathbf{q}) = \rho^*(-\mathbf{q})$ regardless of the approximation used for the defect GF.

### A. Single-defect Green's function

For the single-defect problem, the *exact* GF can be expressed in terms of the $T$ matrix as

$$\mathbf{G}_{\mathbf{k}\mathbf{k}'}(\varepsilon) = \delta_{\mathbf{k},\mathbf{k}'}\mathbf{G}^0_{\mathbf{k}}(\varepsilon) + \mathbf{G}^0_{\mathbf{k}}(\varepsilon)\mathbf{T}_{\mathbf{k}\mathbf{k}'}(\varepsilon)\mathbf{G}^0_{\mathbf{k}'}(\varepsilon), \tag{S5}$$

where the boldface symbols denote matrices in the band ($n$) and spin $s_z$ indices, $\mathbf{k}$ is the electronic wave vector, the matrix $\mathbf{G}^0$ for the *bare* Green's function is diagonal with elements $G^0_{n\mathbf{k}}(\varepsilon) = (\varepsilon - \varepsilon_{n\mathbf{k}} + i\eta)^{-1}$, and the $\mathbf{k}''$ sum is over the BZ. Note that the sum rule $\int d\varepsilon\, \rho(\varepsilon) = N_\mathbf{k} \times N_b$, where $\rho(\varepsilon) = -1/\pi \sum_\mathbf{k} \text{Tr}[\text{Im}\, \mathbf{G}_{\mathbf{k}\mathbf{k}}(\varepsilon)]$, $N_\mathbf{k} = N_k \times N_k$ is the number of $\mathbf{k}$ pointsand $N_b$ is the number of bands, is fulfilled by the bare GF alone, and hence the trace of the correction to the GF in the second term must integrate to zero.

The $T$ matrix describes multiple scattering off a single defect and is given by,

$$\mathbf{T}_{\mathbf{k}\mathbf{k}'}(\varepsilon) = \mathbf{V}_{\mathbf{k}\mathbf{k}'} + \sum_{\mathbf{k}''} \mathbf{V}_{\mathbf{k}\mathbf{k}''}\mathbf{G}^0_{\mathbf{k}''}(\varepsilon)\mathbf{T}_{\mathbf{k}''\mathbf{k}'}(\varepsilon). \tag{S6}$$

The defect matrix elements $\mathbf{V}_{\mathbf{k}\mathbf{k}'}$ are given by the matrix elements of the defect-induced scattering potential with respect to the Bloch functions of the pristine lattice (see below). Note that in contrast to $\mathbf{V}_{\mathbf{k}\mathbf{k}'}$, the $T$ matrix is, in general, not hermitian.

#### 1. Numerical details

To solve for the $T$ matrix in Eq. (S6), we recast it as a matrix equation,

$$[\mathbf{I} - \mathbf{V}\mathbf{G}_0(\varepsilon)]\,\mathbf{T}(\varepsilon) = \mathbf{V}, \tag{S7}$$

where the boldface symbols now denote matrices in the band, spin and $\mathbf{k}$-vector indices, and the Green's function matrix $\mathbf{G}_0$ is diagonal. Rather than solving this equation by direct inversion of the matrix $[\mathbf{I} - \mathbf{V}\mathbf{G}_0(\varepsilon)]$, it is numerically more stable and accurate to regard it as a system of coupled linear equation (one set of coupled equations for each column in $\mathbf{T}$ and $\mathbf{V}$) and solve it with a standard linear solver. This requires one factorization followed by a matrix-vector multiplications and scales as $O(M^3)$ where $M$ denotes the matrix dimension.

The calculations presented in the main manuscript are based on $75 \times 75$ **k**-point samplings of the BZ and include the six lowest spin-orbit split conduction bands. This amounts to a matrix dimension of $M = 6 \times 75^2 = 33750$. With the matrix elements represented as 128-bit complex floating-point numbers, the memory requirement for each of the dense complex matrices in Eq. (S7) becomes $33750^2 \times 128/8 \, \text{bytes} \approx 17$ GBs. To tackle the large matrix dimensions and memory requirements in the solution of the matrix equation (S7), we exploit the automatic openMP multithreading of the LAPACK linear solvers and run the calculations as serial jobs on a multicore platform setting `OMP_NUM_THREADS=$NPROCS` where NPROCS specifies the number of CPUs to be used for multithreading.

## B. Details of the atomistic DFT calculations

All the quantities entering the calculation of the defect GF and the FT-STS spectra, i.e. band structures, defect matrix elements, and phase-factor matrix elements, have been obtained on the above-mentioned **k**, **q**-point BZ grids with the GPAW electronic-structure code[2–4], using DFT-LDA within the projector augmented-wave (PAW) method, a DZP LCAO basis, and including spin-orbit interaction[5]. The implementation will be made available in the GPAW software package, and a full account including details of the PAW specific aspects will be published in a forthcoming paper.

In the following two subsections section, we give a brief overview of: 1. the atomistic DFT-based supercell method for the calculation of the defect matrix elements, and 2. the calculation of the phase-factor matrix elements.

### 1. Defect matrix elements

For nonmagnetic defects of type $i$, we consider a scattering potential of the form

$$\hat{V}_i = V_i(\hat{\mathbf{r}}) \otimes \hat{s}_0 \tag{S8}$$

where $\hat{s}_0$ is the identity operator in spin space. For the present purpose, spin-orbit scattering, which is not included in Eq. (S8), can be safely neglected as the spin-orbit terms are small compared to the main contribution to the scattering potential in Eq. (S8).

The real-space part of the scattering potential is taken as the difference in the crystal potential between the lattice with a defect site and the pristine lattice, i.e.

$$V_i(\mathbf{r}) = V_{\text{dis}}^i(\mathbf{r}) - V_{\text{pris}}(\mathbf{r}). \tag{S9}$$

In practice, this is obtained in a large supercell constructed by repetition of the primitive cell and with the defect site located in the center. Due to periodic boundary conditions in the inplane directions, the supercell must be chosen large enough that defects in neighboring supercells do not interact. A common reference for the two potentials on the right-hand side of Eq. (S9) is ensured by imposing Dirichlet boundary conditions on the cell boundaries in the direction perpendicular to the material plane.

In the basis of the Bloch states and with the spin indices written out explicitly, the defect matrix elements can be expressed as

$$V_{i,\mathbf{kk'}}^{mn}(s,s') = \langle m\mathbf{k}s|\hat{V}_i|n\mathbf{k'}s'\rangle = \sum_{s_z} \langle m\mathbf{k}s; s_z|V_i(\hat{\mathbf{r}})|n\mathbf{k'}s'; s_z\rangle, \tag{S10}$$

with $|\cdot; s_z\rangle$ denoting the $s_z$ component of the wave function.

The numerical evaluation of the defect matrix element in Eq. (S10) is based on an LCAO expansions of the Bloch functions of the pristine lattice, $|m\mathbf{k}s\rangle = \sum_{s_z \nu} c_{m\mathbf{k}s}^{\nu s_z} |\nu \mathbf{k} s_z\rangle$, where $\nu = (\alpha, \mu)$ is a composite index for atomic site ($\alpha$) and orbital index ($\mu$) and

$$|\nu \mathbf{k} s_z\rangle = \frac{1}{\sqrt{N}} \sum_l e^{i\mathbf{k}\cdot\mathbf{R}_l} |\nu \mathbf{R}_l\rangle, \tag{S11}$$

are Bloch expansions of the spin-independent LCAO basis orbitals $|\nu \mathbf{R}_l\rangle$, with $N$ denoting the number of unit cells in the lattice and $\mathbf{R}_l$ is the lattice vector to the $l$'th unit cell.

Inserting in the expression for the matrix element in Eq. (S10), we find

$$V_{i,\mathbf{kk'}}^{mn}(s,s') = \sum_{s_z}\sum_{\nu\nu'}(c_{m\mathbf{k}s}^{\nu s_z})^* c_{n\mathbf{k'}s'}^{\nu' s_z}\langle\nu\mathbf{k}|V_i(\hat{\mathbf{r}})|\nu'\mathbf{k'}\rangle$$

$$= \frac{1}{N}\sum_{s_z}\sum_{\nu\nu'}(c_{m\mathbf{k}s}^{\nu s_z})^* c_{n\mathbf{k'}s'}^{\nu' s_z}\sum_{kl}e^{i(\mathbf{k'}\cdot\mathbf{R}_l - \mathbf{k}\cdot\mathbf{R}_k)}\langle\nu\mathbf{R}_k|V_i(\hat{\mathbf{r}})|\nu'\mathbf{R}_l\rangle, \quad (S12)$$

where the factor of $1/N$ stems from the normalization of the Bloch sums in Eq. (S11) to the lattice area $A$, the last factor in the second line is the LCAO representation of the defect potential $V_i(\mathbf{r})$ in the supercell and the $k, l$ sums are over the lattice cells in the supercell. With this, the defect matrix elements can be calculated for arbitrary $\mathbf{k}, \mathbf{k'} = \mathbf{k} + \mathbf{q}$ values.

Note that the factor of $1/N$ in the defect matrix element, which in the context of Eqs. (S6) and (S3) for the $T$ matrix and FT-STS spectra, respectively, should be associated with the number of $\mathbf{k}$-points, i.e. $N \to N_\mathbf{k}$, ensures that the wave-vector sums in those equations are independent of the BZ sampling.

The matrix elements shown in Fig. 2 of the main manuscript, have a different unit from the one defined in Eq. (S10) above. Using that the lattice area can be written $A = N \times A_{\text{cell}}$ where $A_{\text{cell}}$ is the unit cell area, the matrix element in Eq. (S10) can be expressed as

$$V_{i,\mathbf{kk'}}^{mn}(s,s') = \frac{NA_{\text{cell}}}{A}\bar{V}_{i,\mathbf{kk'}}^{mn}(s,s') \equiv \frac{1}{A}\bar{V}_{i,\mathbf{kk'}}^{mn}(s,s') \quad (S13)$$

where $\bar{V}_i$ has units of the 2D Fourier transform of a scattering potential and is independent on $N$.

### 2. Phase-factor matrix element

The matrix element of the phase factor in Eq. (S3),

$$n_{\mathbf{k},\mathbf{q}}^{mn}(\mathbf{G}) = \langle\psi_{m\mathbf{k}}|e^{-i(\mathbf{q}+\mathbf{G})\cdot\hat{\mathbf{r}}}|\psi_{n\mathbf{k}+\mathbf{q}}\rangle = \langle u_{m\mathbf{k}}|e^{-i\mathbf{G}\cdot\hat{\mathbf{r}}}|u_{n\mathbf{k}+\mathbf{q}}\rangle, \quad (S14)$$

can be reduces to an integral over the primitive unit cell as both the $u_{m\mathbf{k}}$s and $\exp(-i\mathbf{G}\cdot\mathbf{r})$ are cell-periodic functions,

$$n_{\mathbf{k},\mathbf{q}}^{mn}(\mathbf{G}) = N\int_\Omega d\mathbf{r}\, u_{m\mathbf{k}}^*(\mathbf{r})e^{-i\mathbf{G}\cdot\mathbf{r}}u_{n\mathbf{k}+\mathbf{q}}(\mathbf{r}), \quad (S15)$$

where $\Omega$ is the unit-cell volume.

In practice, the matrix elements are evaluated by integrating the first expression in Eq. (S14), i.e. the product of the LCAO Bloch functions and the full phase factor $\exp[-i(\mathbf{q}+\mathbf{G})\cdot\mathbf{r}]$. Note that the factor of $N$ in Eq. (S15) is cancelled by the inverse factor originating from the normalization of the Bloch sums in Eq. (S11).

### C. Calculational details

All DFT calculations presented in the main manuscript have been performed with the electronic structure code GPAW[2–4] within the projector augmented-wave method, using LDA, an LCAO double-zeta polarized basis set, and including spin-orbit interaction[5]. The ground-state densities were obtained using a $21 \times 21$ $\mathbf{k}$-point sampling of the BZ with 7.5 Å vacuum to the cell boundaries in the vertical direction. The defect matrix elements were obtained using a $11 \times 11$ supercell. The phase-factor matrix elements were obtained on a $3 \times 3$ $\mathbf{G}$-point grid with $\mathbf{G}_i = m_i\mathbf{b}_1 + n_i\mathbf{b}_2$, $m_i, n_i = -1, 0, 1$. Finally, the calculation of the $T$ matrix and FT-STS spectra are based on $75 \times 75$ $\mathbf{k}, \mathbf{q}$-point samplings of the BZ with a broadening $\eta = 5$ meV.

### D. FT-STS spectra at the band edge

To support our discussion of the energy dependence of the FT-STS spectra in the main manuscript, we show here in Fig. S2 the spectra for $MoS_2$ at the conduction-band edge. At the conduction-band edge in $MoS_2$, quasiparticle scattering can only take place between the states at the bottom of the $K, K'$ valleys. As a consequence, the FT-STS features in Fig. 3(c)+(d) of the main manuscript are reduced to featureless spots at the $\mathbf{q}$ vectors corresponding to intra- ($\mathbf{q} = \mathbf{0}$) and intervalley scattering. Due to the small spin-orbit splitting in the conduction band of $MoS_2$ ($\sim 3$ meV), spin-conserving intervalley scattering with $\mathbf{q} = \mathbf{K}, \mathbf{K}'$ is possible at the band edge in the presence of a finite linewidth broadening which in our calculations are given by the numerical $\eta$ broadening of the bands. This is indeed the case for the Mo vacancy as shown in Fig. S2. However, the intervalley peak is completely absent for the $S$ vacancy as the intervalley matrix element, according to our symmetry analysis below, in this case vanishes identically between the two high-symmetry $K, K'$ points.

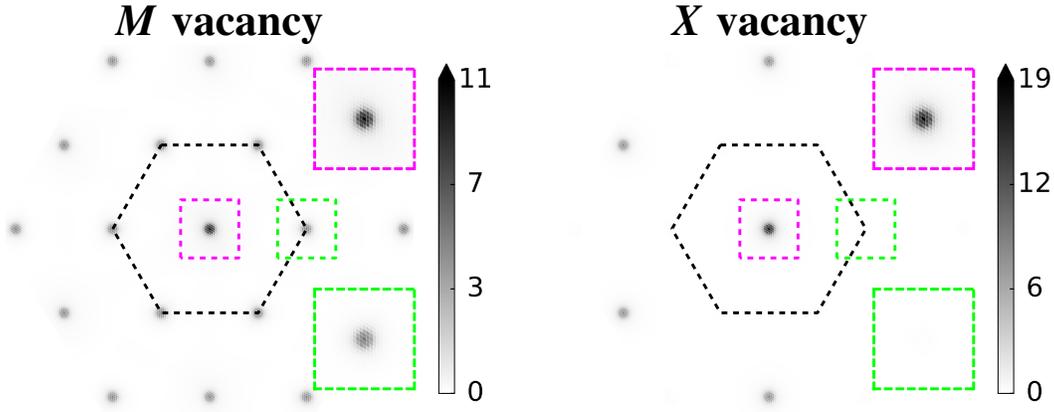

FIG. S2. FT-STS spectra for $M = $ Mo (left) and $X = $ S (right) vacancies in $MoS_2$ at the band edge, $\varepsilon = E - E_c = 0$ meV. The boxes show zooms of the marked regions.

## S2. SYMMETRY ANALYSIS OF THE DEFECT MATRIX ELEMENTS

The selection rules for the matrix elements between states at high-symmetry points in the Brillouin zone can be deduced from the symmetries of the space group which for monolayer $MX_2$ is $D_{3h}$. At the $K, K'$ points, the group of the wave vector is $C_{3h}$ which is formed by the space-group operations $C_3$ (rotation by $\pm 2\pi/3$ around an axis perpendicular to the plane of the monolayer) and $\sigma_h$ (reflection in the horizontal mirror plane defined by the monolayer). Each state at $\mathbf{k} = \pm \mathbf{K}$ can thus be labeled by two quantum numbers which express the phase picked up by the Bloch wavefunction under rotations by $\pm 2\pi/3$ and reflections in the mirror plane, respectively.

When the perturbing defect potential is invariant under one or more of the symmetries forming the group of the wave vector, selection rules for its matrix elements arise. Focusing here on defects with $C_3$ symmetry, the matrix elements in (S10) between the $K, K'$-point Bloch functions labelled by a band ($n = v, c$) and valley ($\tau = \pm 1$) index, can be written

$$\begin{aligned}\langle n\tau|V_i|n\tau'\rangle &= \langle n\tau|C_3^\dagger C_3 V_i C_3^\dagger C_3|n\tau'\rangle \\ &= \langle n\tau|C_3^\dagger V_i C_3|n\tau'\rangle \\ &\equiv \gamma_{i,n}^{\tau\tau'}\langle n\tau|V_i|n\tau'\rangle,\end{aligned} \quad (S16)$$

where $\gamma_{i,n}^{\tau\tau'} = w_{i,n\tau}^* w_{i,n\tau'}$ is given by the product of the phase factors which describe the transformation of the Bloch functions under $C_3$. As we shall see below, the phase factors also depend on the position of the $C_3$ symmetry axis, which is here fixed by the defect type indexed by $i$ ($= M, X$ for $M$ and $X$ centered defects, respectively). From Eq. (S16), it is clear that the matrix element must vanish in case $\gamma_{i,n}^{\tau\tau'} \neq 1$.

The transformation of the symmetry-adapted basis functions[6] defined in the main paper under the $C_3$ symmetry operation can be inferred from their Bloch form,

$$\phi_{n\tau}^K(\mathbf{r}) = \frac{1}{\sqrt{N}} \sum_l e^{i\tau \mathbf{K}\cdot\mathbf{R}_l} \phi_{n\tau}(\mathbf{r} - \mathbf{t}_i - \mathbf{R}_l), \quad (S17)$$

where the sum is over unit cells $l$, $\phi_{n\tau}$ is given by the $d$-orbitals on $M$, and $\mathbf{t}_i$ is the position of the $M$ site in the primitive unit cell with respect to the defect center [see Eq. (S19) below].

Operating on the Bloch functions with $C_3$, we find

$$C_3\phi_{n\tau}^K(\mathbf{r}) = \frac{1}{\sqrt{N}} \sum_l e^{i\tau\mathbf{K}\cdot\mathbf{R}_l} \phi_{n\tau}(C_3^{-1}\mathbf{r} - \mathbf{R}_l - \mathbf{t}_i) \tag{S18a}$$

$$= \frac{1}{\sqrt{N}} e^{-i\tau\mathbf{K}\cdot\mathbf{t}_i} \sum_l e^{i\tau\mathbf{K}\cdot(\mathbf{R}_l+\mathbf{t}_i)} \phi_{n\tau}(C_3^{-1}\left[\mathbf{r} - C_3(\mathbf{R}_l+\mathbf{t}_i)\right]) \tag{S18b}$$

$$= \frac{1}{\sqrt{N}} e^{-i\tau\mathbf{K}\cdot\mathbf{t}_i} \sum_l e^{i\tau C_3\mathbf{K}\cdot C_3(\mathbf{R}_l+\mathbf{t}_i)} \phi_{n\tau}(C_3^{-1}\left[\mathbf{r} - C_3(\mathbf{R}_l+\mathbf{t}_i)\right]) \tag{S18c}$$

$$= \frac{1}{\sqrt{N}} e^{-i\tau\mathbf{K}\cdot\mathbf{t}_i} \sum_l e^{i\tau C_3\mathbf{K}\cdot(\mathbf{R}_l+\mathbf{t}_i)} \phi_{n\tau}(C_3^{-1}\left[\mathbf{r} - \mathbf{R}_l - \mathbf{t}_i\right]) \tag{S18d}$$

$$= \frac{1}{\sqrt{N}} e^{i\tau(C_3\mathbf{K}-\mathbf{K})\cdot\mathbf{t}_i} \sum_l e^{i\tau\mathbf{K}\cdot\mathbf{R}_l} w_{n\tau}\phi_{n\tau}(\mathbf{r} - \mathbf{R}_l - \mathbf{t}_i) \tag{S18e}$$

$$= e^{i\tau(C_3\mathbf{K}-\mathbf{K})\cdot\mathbf{t}_i} w_{n\tau}\phi_{n\tau}^K(\mathbf{r}) \equiv w_{i,\tau}w_{n\tau}\phi_{n\tau}^K(\mathbf{r}) \equiv w_{i,\tau n}\phi_{n\tau}^K(\mathbf{r}). \tag{S18f}$$

Here we have carried out the following steps: (S18a) apply $C_3$; (S18b) insert identity in the form of phase factor; (S18c) inner product invariant under unitary transformation of both vectors; (S18d) summing over $C_3(\mathbf{R}_l+\mathbf{t}_i)$ is the same summing over $\mathbf{R}_l+\mathbf{t}_i$ when the rotation axis is centered on a lattice site; (S18e) $C_3$ is an element in the group of $\mathbf{K} \Rightarrow C_3\mathbf{K}$ and $\mathbf{K}$ are equivalent, and $C_3$ element in the space group $\Rightarrow \phi_{n\tau}(C_3^{-1}[\cdot]) = C_3\phi_{n\tau}(\mathbf{r}) = w_{n\tau}\phi_{n\tau}(\mathbf{r})$, where $w_{n\tau} = e^{2\pi i|m_n|\tau/3}$ originates from the rotation of the orbital around its own center and $m_n$ $(= 0, \pm 2$ for $n = c, v)$ is the magnetic quantum number.

Finally, we evaluate the phase factors defined in Eq. (S18f). In terms of the primitive vectors $\mathbf{a}_{1,2}$ and $\mathbf{b}_{1,2}$ of the direct and reciprocal lattice, respectively, the vectors in (S18f) are given by

$$\mathbf{t}_M = \mathbf{0} \quad \text{or} \quad \mathbf{t}_X = \tfrac{1}{3}\mathbf{a}_1 + \tfrac{1}{3}\mathbf{a}_2, \tag{S19}$$

and

$$\mathbf{K} = -\tfrac{1}{3}\mathbf{b}_1 + \tfrac{1}{3}\mathbf{b}_2 \tag{S20}$$
$$C_3\mathbf{K} = +\tfrac{2}{3}\mathbf{b}_1 + \tfrac{1}{3}\mathbf{b}_2. \tag{S21}$$

For the $C_3$ symmetry axis positioned at the $M$ ($X$) site, we then find $w_{M,\tau} = 1$ ($w_{X,\tau} = e^{2\pi i\tau/3}$). The phase factors from the rotation of the orbitals around their own centers are $w_{v\tau} = e^{4\pi i\tau/3}$ and $w_{c\tau} = 1$.

The $\gamma_{i,n}^{\tau\tau'}$ factor in Eq. (S16) can now be obtained. For the intravalley ($\tau = \tau'$) matrix element, $\gamma_{i,n}^{\tau\tau} = 1$ in all cases implying that the matrix element is finite. On the other hand, for the intervalley ($\tau \neq \tau'$) matrix element we find,

$$\gamma_{M,c}^{\tau\tau'} = 1 \quad , \quad \gamma_{X,c}^{\tau\tau'} = e^{\pm 4\pi i/3}, \tag{S22}$$
$$\gamma_{M,v}^{\tau\tau'} = e^{\pm 8\pi i/3} \quad , \quad \gamma_{X,v}^{\tau\tau'} = e^{\pm 4\pi i/3}, \tag{S23}$$

stating that the intervalley matrix element vanishes identically in all cases except for $M$-centered defects where the matrix element in the conduction band is finite. This is in excellent agreement with our atomistic calculations of the matrix elements in Fig. 2 of the main text.

---